\documentstyle[11pt,epsfig]{article}

\begin{document}
\begin{center}

\title: {\large\bf To ``stabilize'' or not to ``stabilize''?}

\author: {S. Geltman (JILA) and M.
Fotino (MCDB) \\University of Colorado, Boulder, CO 80309-0440,
USA}
\end{center}

 \pagenumbering{arabic}

\begin{abstract}
The controversial question of whether or not an atom will be
completely ionized by a laser pulse of arbitrarily increasing
intensity is discussed.
\end{abstract}

In recent years the world of physics has experienced at least two
highly publicized scientific controversies - ``cold fusion'' and
``the fifth force.''  No doubt there have been many more
smaller-scale disputes that have remained confined to their own
specialized areas.  With this note we would like to bring to the
attention of the wider physics community one such debate that has
been going on for some time within a small subgroup of theoretical
atomic physicists who study the interaction of extremely intense
laser radiation with atoms.

The phenomenon at the center of this debate is referred to as
``stabilization,'' by which is meant the freezing or even decrease
in the photoionization probability of an atom as the laser
intensity is indefinitely increased beyond some ultraintense
value. This of course is a highly counterintuitive expectation,
but a number of theoretical studies using diverse approximations
have obtained results showing various forms of this unusual
behavior [1].

We are on the skeptical side in this debate and feel that those
theoretical studies that have found ``stabilization'' have used
unjustified approximations [2].  The basic objection to the
arguments supporting ``stabilization'' is that the calculations
are not sufficiently rigorous treatments of the problem.  While
perturbative treatments such as the Fermi Golden Rule are
applicable for weak light fields, when the laser intensity becomes
extremely high a rigorous treatment would require a full dynamical
solution.  As is well known from basic nonrelativistic quantum
mechanics, a rigorous treatment of the case of a ground state
hydrogen atom being acted upon by a pulse of intense laser
radiation having electric field ${\bf E}(t)$ would require

\begin{description}
\item[(1)] a solution of the full time-dependent
Schr$\ddot{o}$dinger equation containing the dipole interaction
term ${\mathbf r} \cdot {\mathbf E}(t)$, subject to the initial
condition $\Psi ({\mathbf r},0) = \phi_{1s}(r)$,

\item[(2)]  obtaining the probability amplitude for the bound
electron being ejected into the continuum state $\phi_{\bf
{k}}{(\bf r)}$ as a result of the perturbation, i.e.,
 \begin{equation}
 a_{\bf {k}}(\infty)=\int d{\bf {r}} \phi_{\bf {k}}^* (\bf {r}) \Psi(\bf {r},
 \infty),\nonumber
 \end{equation}
 where ``$\infty$'' may be any time after the laser pulse is over,
 and finally\\

 \item[(3)]  getting the total ionization probability by integrating
 over all continuum states, with the appropriate density-of-states
 factor,
 \begin{equation}
 P_{ion} = \int d{\bf k} (dn/d{\bf k}) |a_{\bf k}(\infty)|^2.
 \end{equation}
\end{description}
 While the above may appear to be a straightforward computational
 task, the fact is that it has not yet been rigorously carried out
 in its entirety for any real atom, not even hydrogen.  Although
 the hydrogenic bound and continuum stationary states are
 analytically well known, there is no known analytic solution for
 the full time-dependent Schr$\ddot{o}$dinger equation for even
 the simplest external oscillatory fields, such as ${\bf E} (t) =
 \bf {E}_o \rm {\sin} \omega t$.  Thus a numerical solution is
 needed, and indeed several numerically accurate solutions for
 $\Psi (\bf {r}, t)$ have been obtained with the help of very
 large-scale computers.  However, steps (2) and (3) have never
 been carried out with full accuracy.  Instead, other
 approximations, which are questionable, have been used to infer
 ``evidence of stabilization'' from the resulting $\Psi (\bf {r},
 \infty)$.  However, for very much simplified atomic models, such as
 a delta function potential in one dimension, it has been possible
 to accurately carry through all the numerical steps above [3].
 The increase in ionization probability as a function of laser
 intensity was found to be not everywhere monotonic, as might be
 expected in the region of tunneling arising from an alternating
 electric field.  However, in the high-field limit the ionization
 probability always goes to 1 and stays there.

 To be credible, ``stabilization'' must be based on a
 \underline{physical} justification for the unexpected reversal of
 ionization probability at ever increasing intensities, rather than
 on a number of disparate calculations using different
 approximation methods that are as yet untested in the ultrahigh
 intensity regime.  When an atom is subject to a static electric
 field that is allowed to increase indefinitely in magnitude, the
 point is reached where the top of the resultant binding potential
 is lowered to the level of the bound state energy and the bound
 electron will spill out, giving rise to the well known phenomenon
 of field emission.  A similar behavior should occur for an
 alternating field, even though the periodic changes in direction
 of the field will possibly reduce the probability of ionization
 from its static field value, and electrons will be ejected in all
 directions.  However, in the case of \underline {fixed frequency},
 what possible physical mechanism could cause an atom to become
 ionized with a \underline {decreasing} probability as (1) the peak electric
 field strength is increasing, and (2) the top of the barrier is
 being continuously lowered, so that the bound electron is
 classically being pulled out to larger displacements from
 the nucleus, and is acquiring larger maximum kinetic energies?
 Until a clear physical answer to this simple question is
 available, doubts about the existence of ``stabilization'' are
 justified in spite of the complex theoretical approximations and
 numerical simulations supporting it that have appeared in the
 literature.

 Some time ago we submitted to Physical Review A a Comment on a
 paper that was published in that journal claiming to have
 observed a case of ``stabilization'' in a laboratory experiment
 [4].  It appeared that there were enough uncertainties in the
 spatial and temporal distributions of the laser fields in the
 vapor cell (three separate lasers were needed to prepare and
 ionize the atoms) to warrant bringing them to the attention of
 the readers.  Included in the Comment were also some of the
 theoretical arguments concerning uncertainties about
 ``stabilization'' as briefly discussed above.  Unfortunately this
 Comment was rejected by Physical Review A through a complex and
 protracted process that appeared to include biased referees'
 reports and unfair editorial handling.  As a result the
 readership of Physical Review A has been deprived of balanced
 views on this issue.  This unpublished Comment is now available
 from the physics archives [5].

 Very recently Dondera et al. [6] have claimed that the ground
 state of atomic hydrogen would undergo ``dynamic stabilization'' by
 not reaching final full ionization with applied laser electric
 fields of up to 80 au.  They infer that they have achieved
 sufficient accuracy in the solution of the time-dependent
 Schr$\ddot{o}$dinger equation over times of up to 100 field
 cycles to allow them to evaluate the ionization probability by
 ``computing the survival probabilities in the discrete states at
 the end of the pulse, and taking the complement of their sum to
 1.''  However, they do not address the obvious question -- how can
 one in practice subtract out the contribution of the infinity of
 Rydberg states that are present?  A more direct and accurate
 procedure would be to project the continuum states directly onto
 the full $\Psi({\bf r},t)$, as indicated in our above equations.

 On the basis of our experience [7] in dealing with the
 time-dependent Schr$\ddot{o}$dinger equation as applied to the
 laser ionization of hydrogen atoms, one cannot accept
 their procedure as being sufficiently accurate.  Their results of
 decreasing ionization probability as a function of increasing
 laser electric field strength in the high-field limit for pulses of
 identical shape and length thus appear to be an artifact of their numerical method.
 It is a misdirection of scientific effort for them to call upon
 experimentalists with ultrahigh-power short-pulse lasers to
 undertake extremely difficult measurements to try to substantiate
 implausible calculational results.

\end{document}